\documentclass[pdflatex,sn-nature]{sn-jnl}
\usepackage{graphicx}%
\usepackage{multirow}%
\usepackage{amsmath,amssymb,amsfonts}%
\usepackage{amsthm}%
\usepackage{mathrsfs}%
\usepackage[title]{appendix}%
\usepackage{xcolor}%
\usepackage{textcomp}%
\usepackage{manyfoot}%
\usepackage{booktabs}%
\usepackage{algorithm}%
\usepackage{algorithmicx}%
\usepackage{algpseudocode}%
\usepackage{listings}%
\usepackage{soul} 
\usepackage{xr} 
\usepackage{geometry}
\theoremstyle{thmstyleone}

\theoremstyle{thmstyletwo}%
\theoremstyle{thmstylethree}%

\begin{document}

\title[Article Title]{Warming-driven rise in soil moisture entropy signals destabilization of the Asian Water Tower}

\author[1]{\fnm{Yiran} \sur{Xie}}
\author*[2,3,4]{\fnm{Teng} \sur{Liu}}\email{teng.liu@tum.de}
\author[5]{\fnm{Xuan} \sur{Ma}}
\author[6]{\fnm{Yingshuo} \sur{Lyu}}
\author[1]{\fnm{Xu} \sur{Wang}}
\author[2]{\fnm{Yatong} \sur{Qian}}
\author[7,8]{\fnm{Yongwen} \sur{Zhang}}
\author*[1,9]{\fnm{Ming} \sur{Wang}}
\email{ wangming@bnu.edu.cn}
\author*[2,10]{\fnm{Xiaosong} \sur{Chen}}\email{chenxs@zju.edu.cn}

\affil[1]{\orgdiv{School of National Safety and Emergency Management}, \orgname{Beijing Normal University}, \city{Beijing}, \postcode{100875}, \country{China}}
\affil[2]{\orgdiv{School of Systems Science and Institute of Nonequilibrium Systems}, \orgname{Beijing Normal University}, \city{Beijing}, \postcode{100875}, \country{China}}
\affil[3]{\orgdiv{Munich Climate Center and Earth System Modelling Group, Department of Aerospace and Geodesy}, \orgname{TUM School of Engineering and Design, Technical University of Munich}, \city{Munich}, \postcode{80333}, \country{Germany}}
\affil[4]{\orgname{Potsdam Institute for Climate Impact Research}, \city{Potsdam}, \postcode{14473}, \country{Germany}}
\affil[5]{\orgname{State Key Laboratory of Disaster Weather Science and Technology, Chinese Academy of Meteorological Sciences}, \city{Beijing}, \postcode{100080}, \country{China}}
\affil[6]{\orgdiv{Department of Earth System Science, Ministry of Education Key Laboratory for Earth System Modeling, Institute for Global Change Studies}, \orgname{Tsinghua University}, \city{Beijing}, \postcode{100084}, \country{China}}
\affil[7]{\orgdiv{Yunnan Key Laboratory of Complex Systems and Brain-Inspired Intelligence}, \orgname{Kunming University of Science and Technology}, \city{Kunming}, \postcode{650500}, \country{China}}
\affil[8]{\orgdiv{Faculty of Science}, \orgname{Kunming University of Science and Technology}, \city{Kunming}, \postcode{650500}, \country{China}}
\affil[9]{\orgname{Joint International Research Laboratory of Catastrophe Simulation and Systemic Risk Governance, Beijing Normal University}, \city{Zhuhai}, \postcode{519087}, \country{China}}
\affil[10]{\orgname{Institute for Advanced Study in Physics and School of Physics, Zhejiang University}, \city{Hangzhou}, \postcode{310058}, \country{China}}

\abstract{
The Tibetan Plateau (TP), known as the ``Asian Water Tower,'' is currently undergoing a rapid wetting trend. While this moisture increase is often viewed as beneficial for water availability, it remains unclear whether the hydrological system itself is becoming more resilient or drifting toward instability. Here, we apply an entropy-based framework to quantify the changing structural organization of the TP's soil moisture system. We show that from 2000 to 2024, regional wetting has driven a long-term decline in entropy, reflecting an increase in system order and stability due to enhanced hydrological buffering capacity. This stability is modulated by the El Ni\~no–Southern Oscillation (ENSO), which regulates regional heterogeneity via a distinct spatial dipole. Crucially, however, CMIP6 climate projections reveal an alarming reversal: future warming triggers a rise in entropy. This transition signals a loss of systemic resilience, characterized by intensified spatial disorder and potential abrupt regime shifts by the mid-century. Our findings suggest that while current wetting provides a stabilizing buffer, continued warming is projected to amplify spatial heterogeneity, thereby destabilizing the Asian Water Tower, with significant risks for downstream water security.
}
\maketitle

\section{Introduction}

Serving as the ``Asian Water Tower," the Tibetan Plateau (TP) is a central regulator of the Asian and global hydrological cycle, storing and redistributing vast quantities of freshwater through its glaciers, snowpack, lakes, permafrost and river networks~\cite{immerzeel2020importance,yao2022imbalance,wang2022contrasting,yao2019recent}. As the world's highest and most expansive plateau, the TP exerts profound influences on atmospheric circulation, monsoon dynamics and ecological stability across Asia. However, its elevated sensitivity to anthropogenic climate change has made it a major area of concern in recent decades~\cite{liu2020land,ren2023environmental}. The TP has been warming at roughly twice the global average rate~\cite{you2021warming}, leading to widespread glacial retreat, permafrost degradation, expanding lakes, and shifts in snowmelt timing~\cite{immerzeel2010climate,wang2022contrasting,yao2019recent}, threatening regional water availability and downstream security~\cite{immerzeel2010climate,immerzeel2020importance,yao2022imbalance,cui2023non}. Beyond these gradual trends, recent evidence suggests that the TP may be approaching the behavior of a climate tipping element~\cite{lu2022warming,liu2023teleconnections,cui2023non,terpstra2025assessment}, a crucial Earth system component prone to abrupt and potentially irreversible transitions under small perturbations~\cite{boers2025destabilization}. Moreover, emerging teleconnections between the TP and other tipping elements, such as the Amazon rainforest, suggest that its destabilization could trigger cascading effects across the broader Earth system, amplifying global climate risks~\cite{liu2023teleconnections,wunderling2023global,wunderling2024climate}.\par

As a critical state variable of the TP’s hydrological system, soil moisture (SM) regulates surface runoff~\cite{ye2023rainfall}, groundwater recharge~\cite{van2024cryosphere}, vegetation dynamics~\cite{cui2024integrating} and carbon fluxes~\cite{humphrey2021soil}, thereby underpinning ecosystem resilience and regional water security. By controlling evapotranspiration (ET) and the partitioning of surface heat fluxes, SM modulates land–atmosphere energy and water exchanges, and influences both regional and remote climate variability and feedbacks~\cite{humphrey2021soil, qiao2023soil}. Specifically, seasonal SM anomalies on the TP modulate the region's thermal forcing, shaping the Asian summer monsoon~\cite{lin2023role,ullah2023empirical} while also triggering teleconnected Rossby waves that drive extreme heatwaves across Europe and East Asia~\cite{gui2024land,jiang2024dry}. However, characterizing the complex spatiotemporal dynamics of SM remains a challenge. Harsh terrain limits ground-based observations, necessitating reliance on satellite products and reanalysis datasets, whose varying performance across the Plateau poses a major challenge for their interpretation~\cite{su2011tibetan,zeng2015evaluation,deng2020responses}. While recent assessments report an overall wetting trend linked to rising precipitation (Pp) and warming-induced alterations in freeze–thaw cycles, pronounced regional differences persist, and the mechanisms driving these spatial contrasts are not well resolved~\cite{hu2024strong,meng2018detecting,deng2020responses,luo2020freeze}. Current understanding is further limited by the complexity of large-scale climate variability. The El Niño–Southern Oscillation (ENSO), for example, is known to disrupt precipitation and cryospheric processes over the Plateau~\cite{shaman2005effect,hu2021impact,hu2022dominant,liu2020large,lei2019extreme,zhu2024extreme,shao2021large}, yet its role in modulating SM dynamics and stability remains poorly quantified.\par
 
A further challenge lies in the lack of a theoretical framework capable of assessing the systemic behavior of SM beyond localized or pixel-based variability. SM dynamics arise from nonlinear interactions among Pp, ET, snowmelt and vegetation feedbacks, and are additionally influenced by stochastic climate variability~\cite{seneviratne2010investigating}. Under external perturbations such as sustained warming, these coupled processes may trigger amplifying mechanisms, such as plant-soil moisture positive feedback~\cite{hu2024plant}, potentially pushing the system toward alternative regimes when stability decreases. Capturing such large-scale behavior requires a metric that goes beyond mean trends to quantify organization, heterogeneity and resilience. Entropy provides precisely such a diagnostic lens. By quantifying the degree of disorder, entropy reveals whether variability is concentrated in a few dominant modes (order) or dispersed across many (disorder), offering a physically interpretable measure of systemic structure and early-warning signals of abrupt transitions~\cite{vranken2015review,tirabassi2023entropy,meng2020complexity}. Here, we leverage recent advances in statistical physics to estimate entropy directly from observational and model data using an eigen microstate approach~\cite{hu2019condensation,liu2022renormalization,ma2024increased,zhang2024eigen,wang2024holistic,xie2025ecosystem}. This method extracts dominant system states to characterize their probability distribution~\cite{Liu2025Phase}, offering a tractable framework for tracking SM disorder under both observed variability and projected climate warming.\par

In this study, we apply this entropy-based diagnostic framework to quantify the spatiotemporal disorder and systemic stability of the SM over the TP. We evaluate the interannual variation and systemic organization using GLDAS-NOAH surface SM data (0–10 cm depth) from 2000 to 2024, selected through comprehensive intercomparison to ensure optimal regional performance. Our analysis reveals a pronounced long-term decline in entropy, indicating increasing spatial organization consistent with the observed wetting trend. Interannual fluctuations in entropy are significantly modulated by ENSO, with El Ni\~no phases generally associated with higher entropy and La Ni\~na with lower entropy, reflecting the imprint of teleconnected climate variability on SM heterogeneity. The Coupled Model Intercomparison Project Phase 6 (CMIP6) projections further suggest a reversal of the recent ordering trend, with entropy rising under warming, accompanied by intensified regional contrasts and, in some models, abrupt mid-century regime shifts. These findings indicate that the TP's SM system may become increasingly unstable, with potential far-reaching implications for the Asian water cycle. \par

\section{Results} 
\subsection{Significant wetting over the TP during 2000--2024}\label{Section 2.1}
Accurately characterizing SM variability over the TP remains a major challenge, primarily due to the scarcity of in situ observations and the region’s complex hydroclimatic conditions. Although several SM datasets, including land surface model outputs (GLDAS-NOAH~\cite{rodell2004global}) and reanalysis datasets (MERRA-2~\cite{reichle2017land}, ERA5-Land~\cite{copernicus2019era5}), are widely employed in global hydrological and climate studies, each exhibits considerable uncertainties over the TP. These discrepancies stem from differences in model structures, parameterizations, and data assimilation strategies~\cite{nearing2018efficiency}. For instance, 
MERRA-2 emphasizes atmospheric consistency but often underrepresents land surface heterogeneity~\cite{reichle2017land}; ERA5-Land provides high-resolution estimates yet exhibits a wet bias in SM climatology in semi-humid regions~\cite{wu2021evaluation}; and GLDAS-NOAH captures SM variability reasonably well but is influenced by uncertainties in the meteorological inputs used to drive the model~\cite{nearing2018efficiency}. Prior studies have highlighted that SM simulations over the TP are particularly sensitive to topography, land cover, and the Pp regime, with model performance exhibiting substantial spatial and temporal variability~\cite{zeng2015evaluation,meng2018detecting}.\par

Considering the inconsistencies across current SM datasets, we first identify the dataset most suitable for our analysis of the spatiotemporal dynamics of SM across the TP. We develop a robust evaluation framework that jointly considers temporal correlation, spectral characteristics, and autocorrelation structure, using in situ observations at 5 cm depth from the Tibet-Obs network, which comprises Maqu, Naqu, Ali, and Shiquanhe subregions (Supplementary Fig. 1). The evaluation incorporates three complementary metrics designed to assess different aspects of temporal fidelity. (1) temporal correlation ($R_T$; see Methods), which quantifies the degree of synchronicity between the evaluated products and site-level measurements.  A high $R_T$ indicates strong agreement in the timing and fluctuation patterns of SM variability. (2) spectral correlation ($R_F$), calculated from the Fourier transforms of evaluated products and observational time series. $R_F$ assesses the preservation of frequency-dependent variability, ensuring the dataset reproduces not only first-order trends but also higher-order dynamical structures embedded in the observed record. (3) autocorrelation sequence similarity ($R_{ACF}$), based on lagged autocorrelation coefficients at lags 1--10 (AC1–AC10). $R_{ACF}$ captures the persistence and memory structure of SM dynamics, with higher similarity indicating a better capability of the evaluated product to replicate the temporal continuity and internal variability of real-world processes.\par

\begin{figure}[htp]
\noindent\includegraphics[width=1\textwidth]{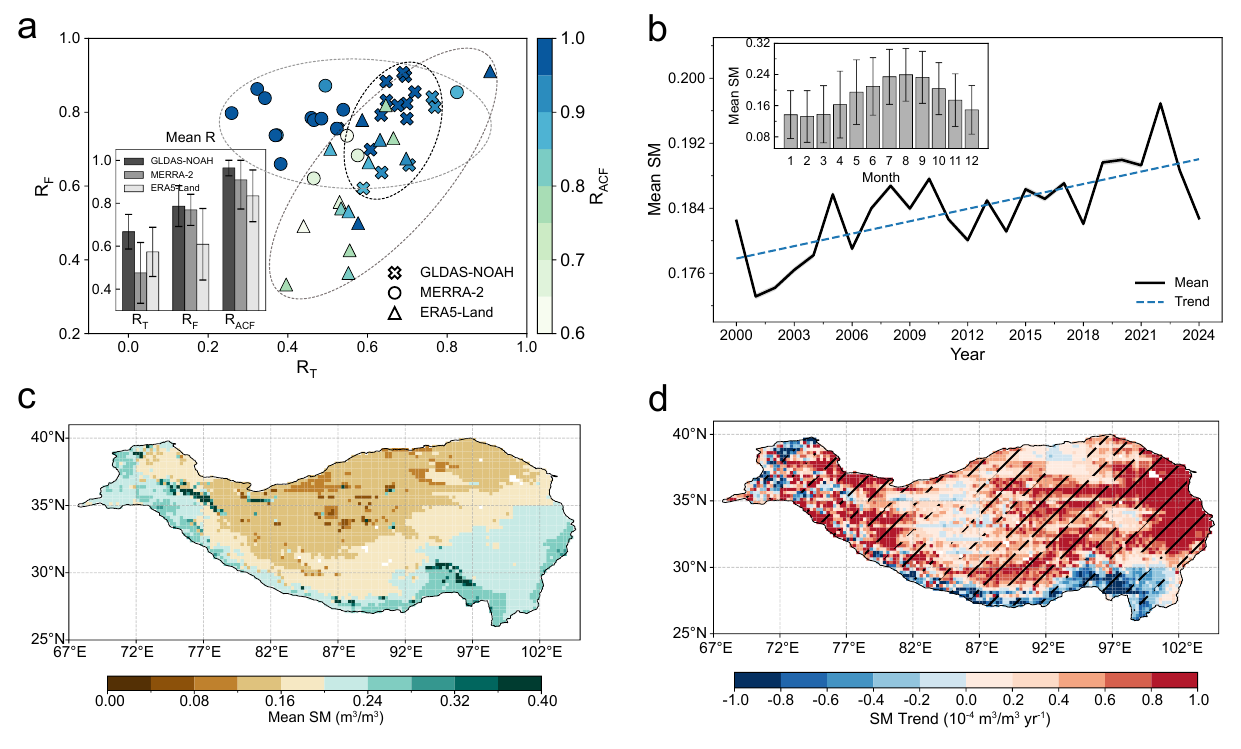}
\caption{\textbf{Spatiotemporal dynamics of SM over the TP.} 
(\textbf{a}) Temporal and spectral correlations ($R_{T}$ and $R_{F}$), with colors indicating the correlation between autocorrelation sequences ($R_{ACF}$), derived from matched SM products. The inset bar chart summarizes the three evaluation metrics, with error bars indicating the standard deviation. 
(\textbf{b}) Temporal evolution of annual mean SM from 2000 to 2024, with shading indicating the standard error. The blue dashed line represents a significant upward trend ($p < 0.001$). The inset bar chart shows the climatological monthly mean SM, with error bars denoting the standard deviation. 
(\textbf{c}) Spatial distribution of mean SM. (\textbf{d}) Spatial patterns of SM trends from 2000 to 2024. Regions with $p<0.05$ are marked with diagonal lines.} \label{fig1}
\end{figure}

We find that GLDAS-NOAH consistently outperforms the other datasets across all evaluation criteria. Specifically, it exhibits the highest $R_T$ with in situ observations and the lowest variability, followed by ERA5-Land, while MERRA-2 performs the worst (Fig.~\ref{fig1}a). This superior performance is consistent across the four examined subregions, demonstrating its stable temporal fidelity at finer spatial scales (Supplementary Fig.~2). Fig.~\ref{fig1}a also shows that GLDAS-NOAH outperforms the others in both mean $R_{F}$ and $R_{ACF}$, indicating its strong ability to capture not only first-order variability but also higher-order dynamical and persistence features of the SM time series. Notably, MERRA-2 outperforms ERA5-Land in $R_{F}$ and $R_{ACF}$, in contrast to the pattern observed for mean $R_{T}$, demonstrating that assessment of dynamical and persistence characteristics yields complementary information beyond conventional first-order statistics. Importantly, the scatterplots in Fig.~\ref{fig1}a show that GLDAS-NOAH displays a more compact distribution across all three metrics, as reflected in the tighter clustering of data points compared to ERA5-Land and MERRA-2. This compact performance distribution underscores the product’s spatial robustness across heterogeneous environments. Although GLDAS-NOAH exhibits a moderate overestimation in certain subregions, such as Ali and Shiquanhe, as indicated by relatively high RMSE and bias values, this issue is less critical when the focus is on relative variability rather than absolute values. In fact, it maintains strong performance in $R_F$ and $R_{ACF}$ across all four subregions, with particularly high consistency in Ali and Shiquanhe (Supplementary Fig.~3). Moreover, at the global scale, GLDAS-NOAH has been shown to capture complex variability exhibiting lower randomness~\cite{kumar2018information}. Based on this comprehensive assessment, we select GLDAS-NOAH as the primary dataset for analyzing the spatiotemporal evolution of SM across the TP.\par

We then analyzed the spatiotemporal variability of SM across the TP from 2000 to 2024 using the GLDAS-NOAH dataset. Spatially, SM exhibits a pronounced southeast-to-northwest decreasing gradient overall, broadly aligning with the regional Pp pattern (Fig.~\ref{fig1}c). This gradient reflects distinct hydroclimatic regimes across the TP: certain high-altitude regions in the southern and northwestern Plateau (e.g., the Himalayas and Hindu Kush) maintain relatively high SM levels due to seasonal snowmelt, whereas most of the arid central and northern Plateau (e.g., the Qaidam Basin and Tanggula Mountains) exhibit markedly lower SM levels. Seasonally, SM follows a distinct annual cycle, increasing from spring to summer and decreasing through autumn and winter. Monthly mean values range from 0.132 $\mathrm{m}^3/\mathrm{m}^3$ in February to 0.239 $\mathrm{m}^3/\mathrm{m}^3$ in August, with an average of 0.184 $\mathrm{m}^3/\mathrm{m}^3$ over the study period (Fig.~\ref{fig1}b). However, this seasonal pattern varies regionally: SM peaks in spring over the Pamir Plateau and Hindu Kush Mountains, whereas in the Himalayas, it remains relatively stable throughout the year (Supplementary Fig. 4). Interannually, SM over the TP exhibits a significant increasing trend, with a linear rate of $+0.0005~\mathrm{m}^3/\mathrm{m}^3$ per year during 2000–2024 (Fig.~\ref{fig1}b). The long-term wetting, statistically significant in all seasons except summer (Supplementary Fig. 4), reflects an intensification of the regional hydrological cycle under a warming climate, primarily driven by enhanced Pp. In summer, however, strong ET associated with high temperatures likely offsets the additional moisture input~\cite{deng2020responses,chen2024doubled}. Nevertheless, substantial regional differences persist, with significant wetting covering approximately 57.5\% of the TP, whereas significant drying is limited to roughly 8.1\%, mainly in the southern Himalayas and parts of the Hengduan Mountains (Fig.~\ref{fig1}d).\par 

\subsection{ENSO-Driven Modulation of SM Entropy on the TP}

A broadly increasing trend in SM across the TP is reshaping regional hydrological dynamics by enhancing soil water storage, altering the partitioning of ET, and modifying runoff generation. These changes affect both the timing and magnitude of water availability across the Plateau and its downstream basins, with important implications for the stability of the regional water cycle. To evaluate soil water stability, we use eigen microstate entropy as a diagnostic of system disorder. This metric quantifies how SM variability is distributed across dominant spatial-temporal modes: lower entropy indicates a state that is governed by a few coherent modes, whereas higher entropy reflects a more disordered state lacking clear organization (see Methods). Applying this framework to the TP's SM reveals a distinct long-term reorganization and clear teleconnections with ENSO events.\par

\begin{figure}[htp]
·\noindent\includegraphics[width=1\textwidth]{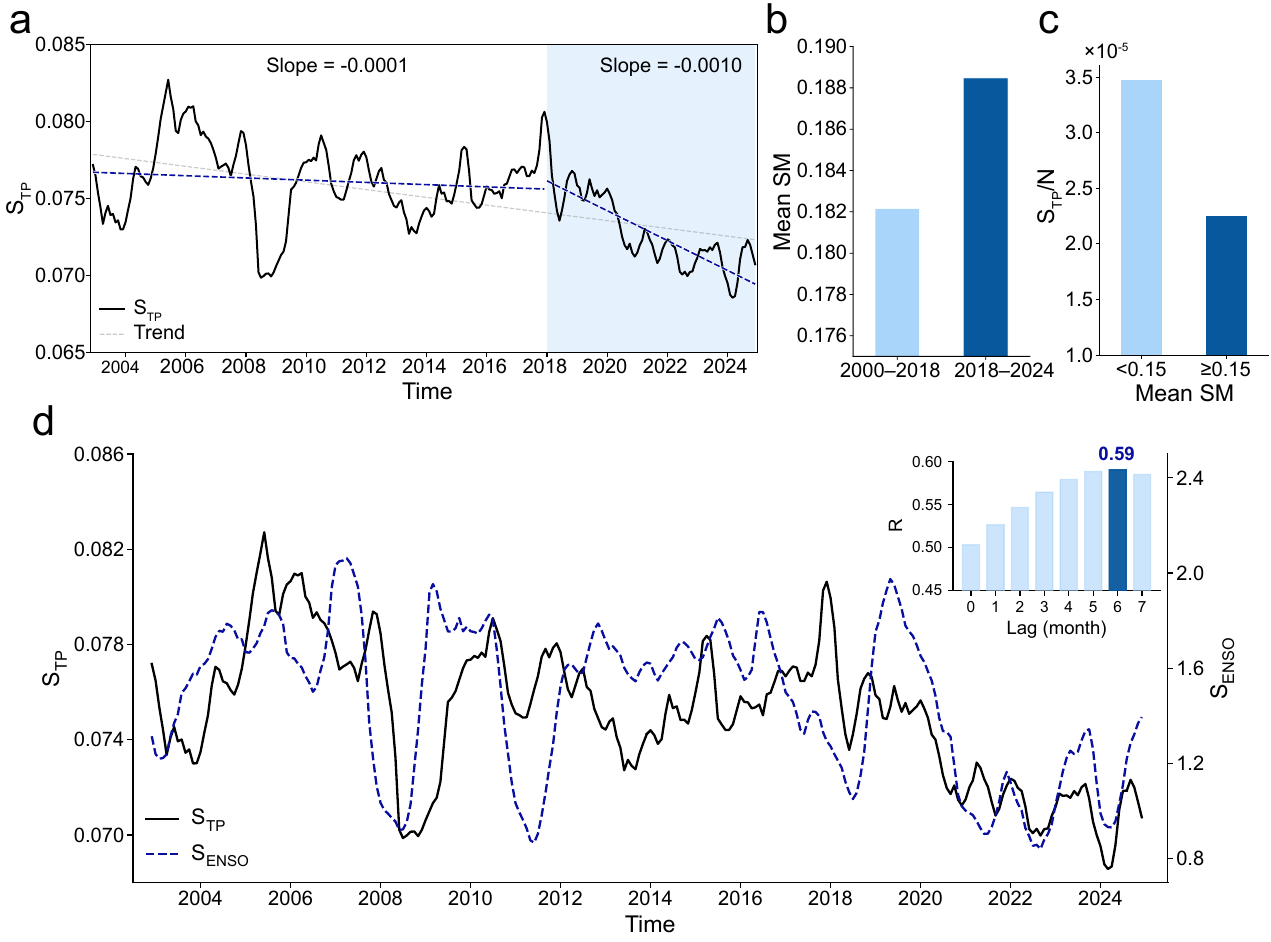}
\caption{\textbf{Evolution of $\boldsymbol{S_\mathrm{TP}}$ and its relationship with ENSO.}
(\textbf{a}) Temporal evolution of $S_\mathrm{TP}$ from 2002–2024 (black), showing a significant decline with a breakpoint in 2018 identified by segmented regression. The decrease is weak before 2018 ($p<0.1$) but accelerates afterwards ($p<0.001$).
(\textbf{b}) Mean SM before and after the 2018 breakpoint, illustrating that wetter conditions correspond to lower disorder.
(\textbf{c}) $S_\mathrm{TP}$ normalized by grid number $N$, showing that locations with higher mean SM ($\ge0.15$)
consistently exhibit lower entropy.
(\textbf{d}) Joint evolution of $S_\mathrm{TP}$ (black) and $S_\mathrm{ENSO}$ (blue), with lag correlations peaking at 0.59 when $S_\mathrm{TP}$ lags $S_\mathrm{ENSO}$ by 6 months, highlighting ENSO modulation of SM entropy on the TP.
} \label{fig2} \end{figure}

The entropy of the TP's SM ($S_{\mathrm{TP}}$) declined significantly during 2000--2024 (Fig.~\ref{fig2}a), with a slope of $-0.000021$ $\mathrm{yr}^{-1}$ ($p<0.001$), indicating an increasingly ordered system under prevailing wetting conditions. A breakpoint in 2018 marks an accelerated decline, with entropy decreasing nearly tenfold faster than before. This shift coincides with higher SM levels after 2018 compared with 2000--2018, supporting the interpretation that wetter states are associated with lower disorder (Fig.~\ref{fig2}b). Spatial patterns reinforce this relationship: locations with higher mean SM (greater than 0.15~$\mathrm{m}^3/\mathrm{m}^3$) consistently exhibit lower entropy (Fig.~\ref{fig2}c), suggesting that ongoing wetting enhances systemic organization across the Plateau. This association arises because wetter soils possess greater hydrological buffering capacity, allowing the system to absorb external perturbations, such as drought events or rapid increases in atmospheric evaporative demand, without large deviations in SM~\cite{fu2024global}. Enhanced buffering stabilizes the temporal evolution of soil moisture, concentrating variability into a narrower set of dominant system states and thus reducing entropy. In contrast, dry soils with limited buffering respond more abruptly to perturbations, generating larger fluctuations and a more dispersed distribution of soil moisture states, which increases entropy.\par

Beyond the long-term ordering, $S_{\mathrm{TP}}$ exhibits pronounced interannual variability that is tightly synchronized with ENSO phases. El Ni\~no years (e.g., 2004, 2015, 2023) correspond to sharp $S_{\mathrm{TP}}$ increases, indicating higher disorder, whereas La Ni\~na years (2007, 2010, 2011, 2017) and the extended 2020--2022 triple La Ni\~na coincide with marked reductions in $S_{\mathrm{TP}}$. To quantify this teleconnection, we calculated the entropy of the ENSO system using the temperature field over the Ni\~no 3.4 region ($S_{\mathrm{ENSO}}$, see Methods). $S_{\mathrm{TP}}$ lags $S_{\mathrm{ENSO}}$ by about six months, yielding a peak correlation of 0.59 ($p<0.001$) (Fig.~\ref{fig2}d), indicating that tropical Pacific variability leaves a delayed and coherent imprint on TP soil moisture disorder. This connection is notable given the large geographical separation between the TP and the tropical Pacific, and the fact that it links entropy derived from temperature anomalies to entropy derived from soil moisture states. The signal is further supported by a significant positive correlation between $S_{\mathrm{TP}}$ and the ENSO index ($R=0.55$, $p<0.01$; Supplementary Fig.~5), and is independently reproduced in CMIP6 historical simulations (Supplementary Fig.~6). When restricted to ENSO-neutral periods, however, the relationship weakens substantially ($R=0.17$, $p=0.05$; Supplementary Fig.~7), indicating that ENSO phases are the primary driver of the observed co-variability. Together, these results demonstrate a robust teleconnection, with ENSO phases exerting a clear modulation of SM entropy across the Plateau.\par

\subsection{ENSO-induced regional heterogeneity governs SM disorder}\label{sec:ENSO_SM}

\begin{figure}[htp]
\noindent\includegraphics[width=1\textwidth]{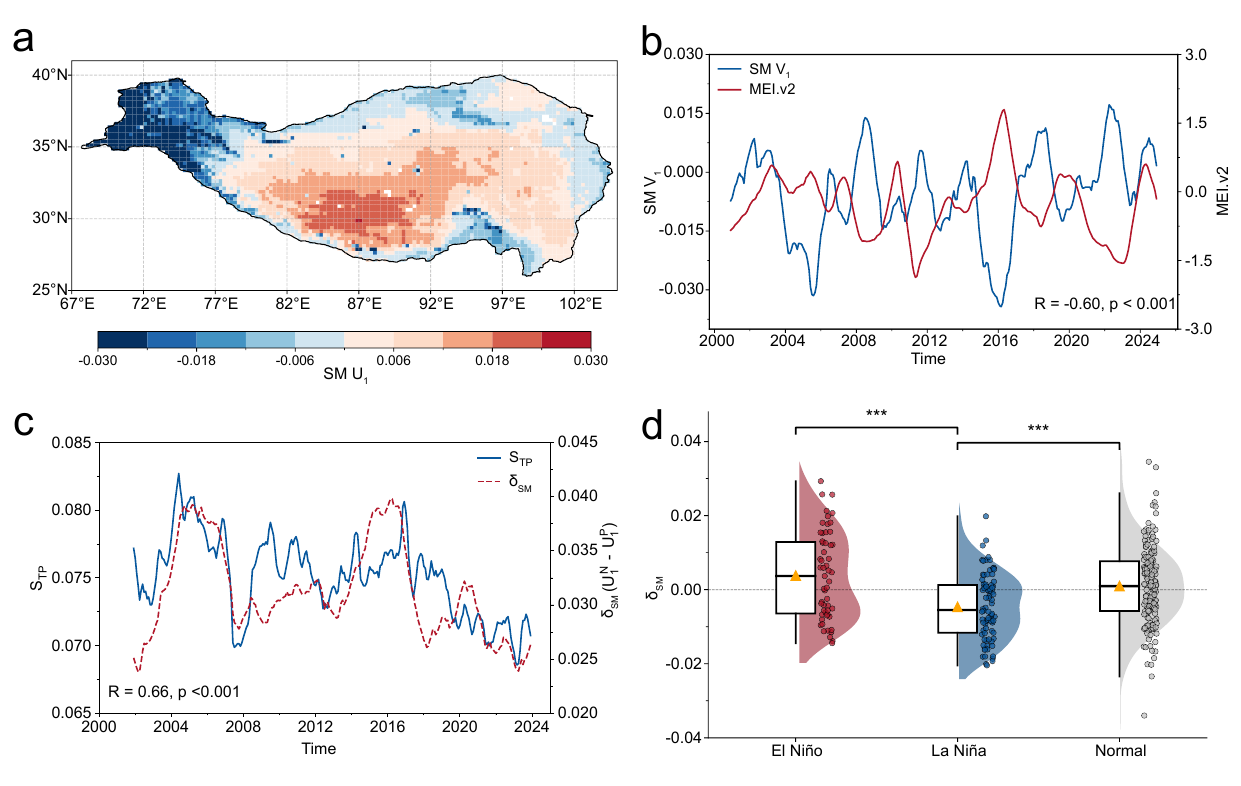}
\caption{\textbf{Relationships between the spatiotemporal patterns of soil moisture EM1, $\boldsymbol{S_\mathrm{TP}}$ and ENSO.} 
(\textbf{a}) Spatial patterns of the soil moisture EM1, highlighting a west–southwest dipole. 
(\textbf{b}) The 12-month smoothed temporal evolution of soil moisture EM1 (SM $V_1$, blue) and ENSO index (MEI.v2, red), showing a strong correlation ($r=-0.6$, $p<0.001$).
(\textbf{c}) Co-evolution of $S_\mathrm{TP}$ (blue) and 3-year running mean SM difference between negative and positive phase regions in SM $U_1$ ($\delta_\mathrm{SM}(U_1^{N} - U_1^{P})$, red), with a correlation of 0.66 ($p < 0.001$), indicating that larger spatial contrasts correspond to higher entropy. 
(\textbf{d}) Violin plots of the two-month running mean SM anomaly difference between positive and negative EM1 phase regions, with individual data points overlaid, during El Ni\~no (red), La Ni\~na (blue), and neutral (gray) periods. Boxplots show the median,  the 25th and 75th percentiles (box edges), 95\% confidence interval (whiskers) and mean values (yellow triangles). Significant differences (***, $p < 0.001$) were identified between El Ni\~{n}o and La Ni\~{n}a, and between La Ni\~{n}a and neutral, based on both one-way ANOVA and two-sided Mann–Whitney $U$-test, whereas no significant difference was found between El Ni\~{n}o and neutral.}
\label{fig3}
\end{figure}

The mechanism underlying ENSO modulation can be understood from the leading eigen microstate of SM (EM1, Fig.~\ref{fig3}a), which explains $\sim35\%$ of total entropy variability (Supplementary Fig. 8). EM1's temporal component $V_1$ (Fig.~\ref{fig3}b), smoothed with a 12-month moving average, shows a significant negative correlation with the Niño 3.4 index ($R = -0.60$, $p < 0.001$) and dominant periodicities of 2–3.8 and 4.4–6 years that align with canonical ENSO cycles (Supplementary Fig. 9). Pixel-wise correlations between the global temperature field and $V_1$ further recover the characteristic ENSO pattern over the tropical Pacific (Supplementary Fig. 10), reinforcing the close relationship between ENSO and EM1. EM1's spatial component $U_1$ exhibits a pronounced dipole structure, with opposite phases in the western and southwestern TP (Fig.~\ref{fig3}a). This dipole highlights the mechanism of ENSO influence: by enhancing or suppressing spatial heterogeneity, ENSO directly modulates SM entropy over the TP. \par

A series of connection analyses support this mechanism. As shown in Fig.~\ref{fig3}c, $S_\mathrm{TP}$ correlates strongly with the SM difference between the positive- and negative-phase regions of $U_1$ ($R = 0.66$, $p < 0.001$), demonstrating that greater spatial contrasts correspond to higher entropy, whereas more coherent conditions correspond to lower entropy. We confirm that ENSO intensity is positively associated with this interregional difference ($R=0.65$, $p<0.001$; Supplementary Fig.~11a): El Ni\~no events amplify regional disparities, while La Ni\~na suppresses them. This spatial asymmetry is further illustrated in Fig.~\ref{fig3}d, where El Ni\~no years typically feature positive SM anomalies between the dipole regions--wetter in the negative-phase region and drier in the positive-phase region. By contrast, La Ni\~na years show the opposite pattern, while neutral years exhibit minimal contrasts. The monthly interregional difference also closely tracks the ENSO index (Supplementary Fig.~11b, $R=0.37$, $p<0.001$), underscoring the robust linkage between ENSO variability and regional hydroclimatic contrasts.\par

\begin{figure}[htp]
\noindent\includegraphics[width=1\textwidth]{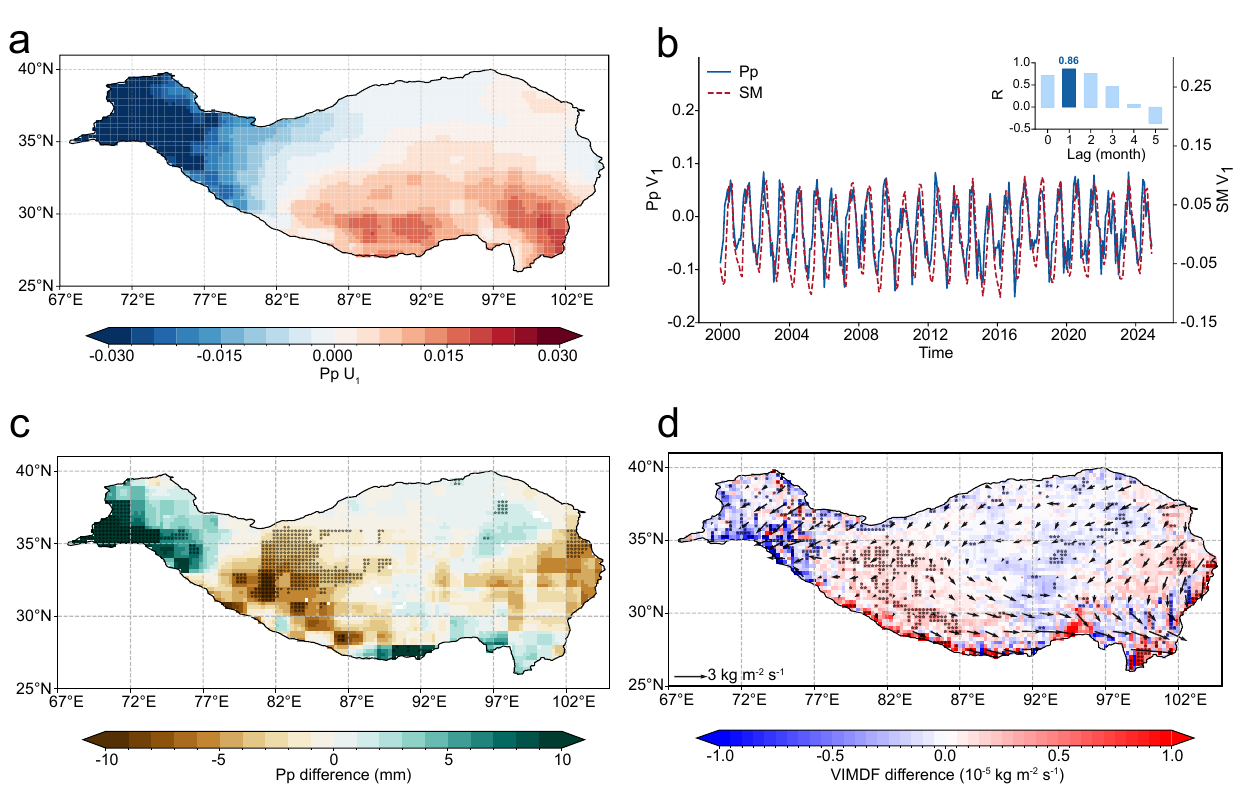}
\caption{\textbf{Spatiotemporal patterns of precipitation EM1 and its relationships with SM and ENSO.} 
(\textbf{a}) Spatial patterns of the precipitation EM1 (Pp $U_1$). 
(\textbf{b}) Temporal evolution of precipitation EM1 (Pp $V_1$, blue) and soil moisture EM1 (SM $V_1$, red), with a strong lagged correlation peaking when SM lags Pp by one month, indicating a rapid SM response to precipitation variability in the TP.
(\textbf{c}) Mean monthly Pp anomaly difference between ENSO phases (El Ni\~no $-$ La Ni\~na).
(\textbf{d}) Difference in vertically integrated water vapour flux anomaly (vectors) and its divergence anomaly (VIMDF, shaded) between ENSO phases (El Ni\~no - La Ni\~na). Asterisks denote regions with statistically significant differences based on a two-tailed Welch’s \textit{t}-test ($p<0.1$).}

\label{fig4}
\end{figure}

The physical mechanisms underlying ENSO-driven interregional contrasts were further examined. Distinct moisture regimes characterize the dipole: the western TP exhibits a significant negative phase, with SM peaking in May due to snowfall, spring melt, and low ET, whereas the southwestern TP shows a significant positive phase, with SM peaking in August under the influence of monsoonal rainfall (Supplementary Figs.~9a, 12). Strong spatial coherence between SM and Pp across the $U_1$ field ($R=0.78$, $p<0.001$), along with temporal synchrony peaking at a one-month lag ($R=0.86$, $p<0.001$), demonstrates that Pp is the primary driver of SM variations (Figs.~\ref{fig4}a, b). The interregional SM contrast, closely tracking the Pp difference (Supplementary Fig.~13a, $R=0.70$, $p<0.001$), provides further evidence for this relationship. Moreover, the annual Pp differences correlate significantly with the ENSO index, suggesting that ENSO strongly modulates TP precipitation patterns (Supplementary Fig.~13b, $R=0.54$, $p<0.01$). Spatially, El Ni\~no events tend to increase Pp over the western TP and reduce it over the southwestern TP, with La Ni\~na generally producing the opposite pattern (Fig.~\ref{fig4}c). ENSO-induced changes in moisture transport likely underpin this mechanism. The composite anomalies show that during El Ni\~no events, southward moisture-flux anomalies and moisture convergence occur over the western TP, while westward and southward transport anomalies and moisture divergence are found over the southwestern Plateau (Fig.~\ref{fig4}d). This pattern is consistent with previous findings that during El Ni\~no events, the weakened and eastward-shifted Walker circulation induces descending motion over the eastern Indian Ocean and the Maritime Continent, which in turn excites a pair of anomalous anticyclones straddling the equator in the tropical Indian Ocean~\cite{wang2002atmospheric}. This configuration strengthens cyclonic anomalies over the Bay of Bengal and the South China Sea, weakening eastward and southward moisture transport over the southwestern TP. Simultaneously, it forms an anticyclonic ridge over the tropical north Indian Ocean that enhances northward moisture transport toward the western TP.~\cite{hu2021impact,liu2020large}.\par

By quantifying the system’s disorder, entropy offers a novel perspective on ENSO–SM teleconnections over the TP, revealing aspects that are not captured by traditional measures of ENSO intensity. The observed six-month lag in this teleconnection likely reflects the cumulative response of the hydrological system. Specifically, surface SM responds to Pp with a delay of approximately one month. In high-altitude regions where Pp falls as snow, snowmelt further postpones its increase~\cite{meng2018detecting,deng2020responses}. Seasonal freeze-thaw cycles in the TP’s permafrost regions impose an additional delay on the adjustment of surface SM to climate anomalies~\cite{luo2020freeze}. Another potential explanation involves the gradual integration of spatially asynchronous seasonal hydrological anomalies. For instance, during El Ni\~no phases, reduced summer rainfall in the southwestern TP and enhanced winter snowfall in the western region are incorporated over time, further delaying the adjustment of $S_\mathrm{TP}$. In conclusion, eigen microstate entropy offers a system-level perspective on the non-linear, cumulative mechanisms of climate-hydrology coupling, informing climate prediction and water resource management on the TP.\par

\subsection{Projected drying and increasing disorder of the SM system over the TP}

\begin{figure}[htp]
\noindent\includegraphics[width=1\textwidth]{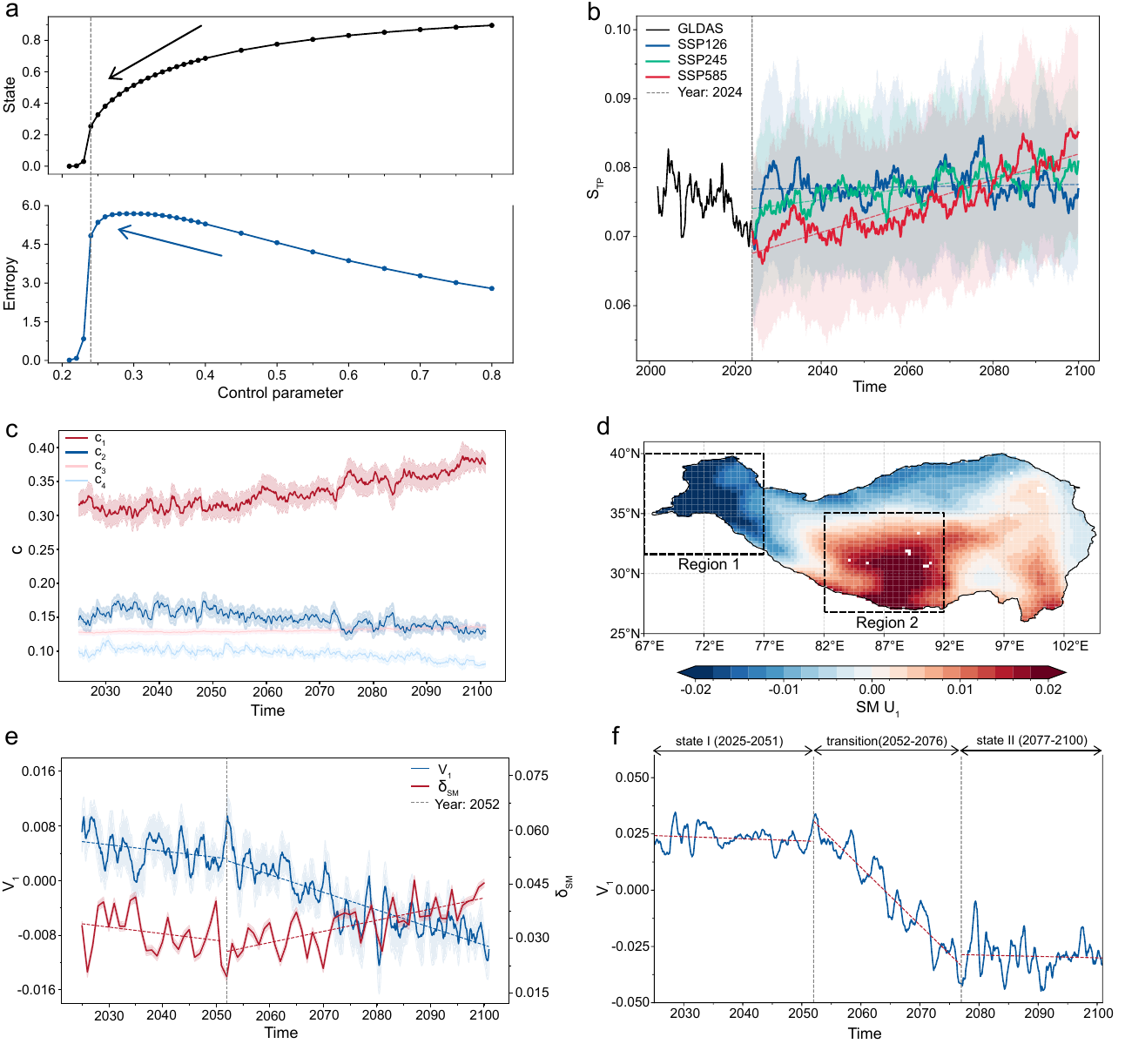}
\caption{\textbf{Projected evolution of SM over the TP.} 
(\textbf{a}) LFCA system dynamics showing a collapse of the system state (black) below a critical threshold (grey), accompanied by rising entropy (blue), which signals increasing disorder before an abrupt change.
(\textbf{b}) Observed (2000--2024, black) and projected (2025--2100) $S_\mathrm{TP}$ from CMIP6 models under SSP126 (blue), SSP245 (green), and SSP585 (red), all exhibiting significant upward trends ($p < 0.001$).
(\textbf{c}) Temporal evolution of mean contributions of the four dominant eigen microstates to $S_\mathrm{TP}$ under SSP585 scenarios.
(\textbf{d}) Ensemble-mean spatial pattern of soil moisture EM1 under SSP585 ($U_1$). Region 1 (western TP, significant negative phase) and Region 2 (southwestern TP, significant positive phase) are marked by blue and red rectangles, respectively.
(\textbf{e}) Ensemble-mean temporal evolution of soil moisture EM1 ($V_1$, blue) under SSP585, with a breakpoint in 2052 (grey) identified by segmented linear regression and a significant post-breakpoint decline ($p<0.001$). The red line denotes the annual mean SM difference between the two regions in (\textbf{d}). No significant trend is observed before 2052, while a significant decline occurs thereafter ($p<0.001$).
(\textbf{f}) Temporal evolution of soil moisture EM1 for the IPSL-CM6A-LR model under SSP585 ($V_1$, blue line), illustrating a sharp mid-century transition followed by stabilization. Three distinct stages are identified: a relatively stable state during 2025–2051, a rapid decline during the transition period (2052–2076, $p<0.001$), and a stable low state during 2077–2100.} 
\label{fig5}
\end{figure}

Rapid warming over the TP is expected to elicit nonlinear responses in the SM system, potentially triggering abrupt regional shifts under future climate scenarios~\cite{liu2023teleconnections,terpstra2025assessment,zaitchik2023wetting,qiao2023soil}. Projected increases in ENSO frequency and amplitude~\cite{cai2021changing} may further destabilize SM patterns through the teleconnections described above. Moreover, Soil moisture--atmosphere coupling over the TP may propagate via Rossby wave pathways, potentially modulating remote climate extremes, including heatwaves over Europe and East Asia~\cite{jiang2024dry,qiao2023soil}, highlighting the importance of assessing projected TP soil moisture changes under warming scenarios. In complex systems, rising entropy has been proposed as an early-warning signal of reduced stability, preceding abrupt transitions across diverse contexts, from theoretical models to biological processes~\cite{tirabassi2023entropy,vranken2015review}. To illustrate this, we consider a Local Facilitation Cellular Automaton (LFCA) model~\cite{kefi2007local}, which undergoes a bifurcation-like shift when a control parameter changes. Unlike the one-dimensional Ornstein-Uhlenbeck process often used in climate tipping point studies~\cite{smith2023reliability}, the LFCA model is inherently high-dimensional, allowing it to capture interactions among multiple components. This property makes it more representative of real-world complexity and well-suited to eco-hydrological processes such as vegetation-wetness interactions~\cite{tirabassi2023entropy}. As shown in Fig.~\ref{fig5}a, system state declines as the control parameter weakens, while the entropy of system state ($S_\mathrm{LFCA}$) rises in advance of the threshold, signaling the approach of an abrupt transition. \par

We therefore turn to CMIP6 projections to assess how the TP's SM entropy evolves under future warming. We analyzed monthly SM at 10 cm depth from the 12 CMIP6 models under SSP126, SSP245, and SSP585, selecting models that best reproduced historical $S_\mathrm{TP}$ trajectories from the GLDAS-NOAH dataset during 2000--2024. The ensemble mean of $S_\mathrm{TP}$ from these models shows strong agreement with observations ($R = 0.57$, $p < 0.0001$; Supplementary Fig.~14), indicating that the selected ensemble captures the observed variability and is suitable for assessing future changes. Across scenarios, $S_\mathrm{TP}$ reveals a pronounced upward trend with warming (Fig.~\ref{fig5}b), indicating a progressive rise in disorder and a potential early warning signal of stability loss in the TP's SM system. The projected rise in entropy corresponds to anticipated declines in annual SM anomalies through 2100, particularly under SSP585 (Supplementary Fig.~15). The direction of this co-variation is consistent with our observational finding that, during 2000–2024, decreases in $S_\mathrm{TP}$ were accompanied by increases in SM, thereby reinforcing the negative relationship between moisture availability and system entropy. Meanwhile, the contribution of the dominant dipole mode, which primarily drives the increase in $S_\mathrm{TP}$ (Fig.~\ref{fig5}c), intensifies, particularly under higher emission scenarios (Supplementary Fig.~16). This amplification highlights the enhanced SM heterogeneity between Region 1 (western TP, significant negative phase) and Region 2 (southwestern TP, significant positive phase), underscoring the importance of regional contrasts in shaping future instability (Fig.~\ref{fig5}d; Supplementary Fig.~17).\par

Consistent with the increase in $S_\mathrm{TP}$, additional evidence of potential abrupt shifts in the TP's SM emerges from the temporal evolution of the dominant dipole mode ($V_1$) under the SSP585 scenario. Segmented regression of $V_1$ reveals a significant breakpoint around 2052 ($p < 0.001$), followed by an accelerated decline after 2052 at more than twice the pre-2052 rate (Figs.~\ref{fig5}e). Notably, $S_\mathrm{TP}$ also shows a significant change at the same time (Supplementary Fig. 18). Coupled with its spatial pattern $U_1$ (Figs.~\ref{fig5}d), the post-2052 divergence between Region 1 and Region 2 drives a marked increase in regional heterogeneity, whereas differences prior to 2052 were negligible, as indicated by the red line in Fig.~\ref{fig5}e. At the model level, IPSL-CM6A-LR exhibits a distinct regime shift (Fig.~\ref{fig5}f): an initial modest decline during 2025--2051 (Stage I), a pronounced transition phase with rapid decrease during 2052--2076, and stabilization into a new state after 2077 (Stage II). Compared with Stage I, Stage II is characterized by widespread SM reductions across most of the Plateau, with the northeast remaining comparatively wetter (Supplementary Fig. 19). Notably, similar abrupt transitions have been identified in this model for snow-related variables under warming levels of 0.53--2.61~K~\cite{terpstra2025assessment}, underscoring the broader susceptibility of TP hydroclimate components to threshold-like responses.\par

The abrupt SM decline after mid-century may reflect the emergence of interacting positive feedbacks that amplify drying once key thresholds are exceeded~\cite{boers2025destabilization}. Reduced SM can weaken evapotranspiration and reduce cloud formation, enhancing short-wave radiation and shifting the surface energy balance toward sensible heating, thereby intensifying land–atmosphere warming~\cite{qiao2023soil}. At the same time, progressive drying may favor stress-tolerant, shallow-rooted vegetation with lower water-use efficiency and reduced soil water retention, weakening the system’s buffering capacity~\cite{hu2024plant}. The convergence of these fast atmospheric and slower ecological adjustments could render the SM system increasingly sensitive to small climatic perturbations, potentially giving rise to threshold-like behavior consistent with the modeled transition.

\section{Discussion and Conclusion}

Pronounced warming across the TP is driving substantial changes in regional hydrology, raising concerns over the stability of the “Asian Water Tower”~\cite{yao2022imbalance}. Beyond local impacts, the TP is embedded in a network of teleconnections among potential climate tipping elements, including ENSO, the South Asian monsoon, and even the Amazon rainforest~\cite{lenton2008tipping,liu2023teleconnections}. Destabilization in one component can trigger cascading responses in others, thereby amplifying systemic risks~\cite{dekker2018cascading}. Our results provide new evidence for such teleconnections: SM entropy on the TP is sensitive not only to regional wetting and drying but also to ENSO variability, which modulates spatial heterogeneity through a dipole pattern. This finding reinforces the view that the TP interacts dynamically with other global climate components, implying that instabilities on the Plateau may exert far-reaching influences through teleconnections \cite{you2021warming,wang2022contrasting,liu2023teleconnections}.\par

Future projections reveal a reversal of the recent ordering trend, with $S_\mathrm{TP}$ rising in parallel with projected SM drying. The consistency between historical observations (lower entropy during wet periods) and projections (higher entropy during drying) underscores a robust negative relationship between water availability and systemic disorder. Crucially, the projected amplification of the dipole mode suggests that intensifying spatial contrasts between the western and southwestern TP will increasingly shape future hydroclimatic instability. Such intensified heterogeneity may amplify extreme events. The July 2022 Qiangtang Plateau heatwave, for instance, was exacerbated by negative SM anomalies through strong land–atmosphere coupling~\cite{gui2024land}. Among the models analyzed, IPSL-CM6A-LR exhibits a distinct mid-century regime shift, with an abrupt transition followed by stabilization into a new state of widespread drying. This, together with the projected rise in disorder, resembles early-warning signals of critical transitions observed in a conceptual model, highlighting the potential for abrupt, nonlinear shifts in TP hydroclimate~\cite{terpstra2025assessment}.\par

Rising entropy and enhanced spatial heterogeneity across the TP suggest that water availability will become increasingly uneven. This may have cascading implications for major downstream river basins in South and East Asia, which collectively support nearly two billion people. Negative spring SM anomalies in the western TP, for instance, have been linked to extreme summer Pp in the Yangtze River Basin, intensifying downstream flood risk~\cite{zhu2023diagnosing}. At the same time, SM-driven circulation anomalies across the Plateau, reinforced by strong land–atmosphere feedbacks, are identified as key drivers of concurrent extremes, including East Asian heatwaves and South Asian floods~\cite{he2023extremely}. Increasing TP instability may therefore pose significant threats to agricultural production, hydropower generation, and water security across densely populated regions~\cite{yao2022imbalance, cui2023non}. The emergence of regime-shift behavior under high-emission scenarios further suggests that warming could undermine the long-term predictability of water supply and complicate adaptation planning.\par

Despite these advances, several key uncertainties remain. In this study, entropy is not interpreted as a direct measure of stability, but rather as a diagnostic of the system’s degree of structural organization. Rising entropy reflects a progressive dispersion of variability across multiple competing modes and is often interpreted as a loss of coherent organization associated with reduced resilience and heightened sensitivity to perturbations in complex, high-dimensional systems.~\cite{de2014measure,tirabassi2023entropy,Liu2025Phase}. Although we employed rigorous dataset screening, observational constraints remain due to sparse in-situ networks and satellite retrieval biases. The coarse spatial resolution of SM products may also underestimate localized heterogeneity, particularly in steep or high-relief regions~\cite{zeng2015evaluation,su2011tibetan}. Model simulations also exhibit known deficiencies in representing SM processes in high-altitude and semi-arid environments. Moreover, while Pp is a primary driver, the contribution of snow accumulation, glacier melt, and surface runoff to entropy dynamics remains insufficiently assessed. Finally, evidence of abrupt regime shifts is currently confined to a subset of models, leaving open the question of whether these signals represent genuine thresholds or model-specific artifacts.\par

Collectively, these findings underscore the urgency of sustained monitoring of the TP's SM and improved representation of land--climate feedbacks in Earth system models, particularly with respect to cryospheric processes and teleconnections. More broadly, our work highlights the value of entropy-based diagnostics for detecting early-warning signals of hydrological instability. Applying such frameworks to other climate-sensitive regions may provide a powerful tool for anticipating tipping-like behavior in the terrestrial water cycle and its cascading impacts on society.\par

\clearpage

\clearpage

\clearpage

\clearpage

\clearpage

\clearpage

\section{Methods}\label{Method}

\subsection{Data}

\subsubsection{SM Datasets}

This study integrates three complementary sources of SM data: (1) in-situ observations, (2) reanalysis products, and (3) land surface model outputs. In-situ SM observations were obtained from the Tibetan Plateau Observatory (Tibet-Obs)~\cite{zhang2020status}, a ground-based monitoring network that spans major climatic regimes: cold arid (Ali, 4 sites; Shiquanhe, 20 sites), semi-arid (Naqu, 11 sites), and humid (Maqu, 26 sites) climates (Supplementary Fig.~1). Surface SM was measured at 5 cm depth and 15-min resolution during 2009--2019. To ensure consistency with gridded products, the observations were aggregated to monthly means and mapped to a $0.25^\circ \times 0.25^\circ$ grid.

Reanalysis and land surface model datasets include the Modern-Era Retrospective Analysis for Research and Applications Version 2 (MERRA-2), ERA5-Land reanalysis dataset produced by the European Center for Medium-Range Weather Forecasts, and Global Land Data Assimilation System Noah Land Surface Model version 2.1 (GLDAS-NOAH). MERRA-2 provides monthly SM at $0.5^\circ \times 0.625^\circ$ resolution across three soil layers: $0$--$5~\mathrm{cm}$, $5$--$100~\mathrm{cm}$, and $100$--$289~\mathrm{cm}$. ERA5-Land offers monthly SM at $0.1^\circ$ resolution across four layers: $0$--$7~\mathrm{cm}$, $7$--$28~\mathrm{cm}$, $28$--$100~\mathrm{cm}$, and $100$--$289~\mathrm{cm}$. GLDAS-NOAH provides monthly SM at $0.25^\circ$ resolution with four layers: $0$--$10~\mathrm{cm}$, $10$--$40~\mathrm{cm}$, $40$--$100~\mathrm{cm}$, and $100$--$200~\mathrm{cm}$. This study focuses on the top soil layer for each product: $0$--$5~\mathrm{cm}$ (MERRA-2), $0$--$7~\mathrm{cm}$ (ERA5-Land), $0$--$10~\mathrm{cm}$ (GLDAS-NOAH), over the common analysis period 2000-2024. All datasets were resampled to a uniform $0.25^\circ$ resolution, and GLDAS-NOAH SM values were converted from mass ($\mathrm{kg~m^{-2}}$) to volumetric units ($\mathrm{m^3~m^{-3}}$) for inter-dataset consistency.\par

\subsubsection{Meteorological data}
This study employs monthly meteorological variables including precipitation (Pp) and evapotranspiration (ET) from the GLDAS-NOAH dataset provided at $0.25^\circ \times 0.25^\circ$ resolution. Near-surface (1000-hPa) air temperature from ERA5 was used to assess the spatial covariability between ENSO signals and SM $V_1$. The monthly total column vertically integrated water vapour flux and its divergence from ERA5 were used to investigate ENSO-related moisture transport influencing Pp over the TP. ENSO variability was quantified using the Multivariate ENSO Index Version 2 (MEI.v2), which combines sea level pressure, sea surface temperature, low-level winds, and outgoing longwave radiation to provide a physically consistent representation of coupled ocean–atmosphere dynamics~\cite{zhang2019towards}. Since MEI.v2 does not define standardized thresholds for the onset or classification of ENSO events, the Oceanic Ni\~no Index (ONI) was additionally used to identify and categorize ENSO phases. El Ni\~no (La Ni\~na) events are defined as periods when the ONI exceeds $+0.5~^{\circ}\mathrm{C}$ (falls below $-0.5~^{\circ}\mathrm{C}$) for at least five consecutive months(see Supplementary Fig.~20).

\subsubsection{CMIP6 model simulations}
To investigate future changes in SM entropy over the TP, we analyzed monthly soil moisture in the upper soil layer (mrsos) from 12 CMIP6 climate models (see Supplementary Table~1). Models were selected according to their ability to reproduce the dynamical characteristics of $S_\mathrm{EM}$ as evaluated against GLDAS-NOAH datasets. To ensure inter-model comparability, all outputs were resampled to a common $0.25^\circ \times 0.25^\circ$ grid using bilinear interpolation and converted from mass $\mathrm{kg~m^{-2}}$ to volumetric units ($\mathrm{m^3~m^{-3}}$). Future projections were examined for 2025--2100 under three Shared Socioeconomic Pathway (SSP) scenarios: SSP126, SSP245, and SSP585. \par

\subsection{Evaluation of SM Data Applicability}
To ensure that entropy estimation and subsequent dynamical analyses are grounded in physically meaningful SM behavior, we evaluated the applicability of three commonly used SM products (GLDAS-NOAH, MERRA-2, and ERA5-Land) over the TP. Rather than focusing on conventional accuracy, the goal is to assess whether these datasets retain the temporal variability and memory characteristics that underpin eigen microstate decomposition and teleconnection analyses.\par

Two complementary aspects were examined. First, absolute agreement was assessed using three metrics that reflect consistency in mean states and overall temporal variability between gridded and in-situ observations: the temporal Pearson correlation coefficient ($R_T$), root mean square error (RMSE), and bias (Bias). Together, they indicate whether the products reproduce the magnitude and basic temporal behavior of observed SM, thereby preventing systematic offsets from propagating into entropy calculations. Second, dynamical similarity was evaluated using metrics that characterize temporal structure. The spectral Pearson correlation coefficient ($R_F$) measures the agreement in frequency-dependent variability, ensuring the preservation of dominant oscillatory components contributing to eigen microstate states. The autocorrelation structure correlation ($R_{ACF}$) quantifies similarity in short- to medium-term memory by comparing autocorrelation functions over the first 10 lags, representative of month-scale persistence in the TP's SM. These properties are critical because the entropy framework is sensitive to the distribution and persistence of spatiotemporal fluctuations.\par

Together, these five metrics provide a targeted evaluation of whether SM products maintain the key statistical properties required for reliable entropy estimation and dynamical interpretation. Full definitions and calculation formulas are summarized in Supplementary Table~2.

\subsection{Eigen Microstates Entropy}

To quantify the degree of disorder of TP soil moisture system, we adapt the eigen microstates theory (EMT)~\cite{hu2019condensation,sun2021eigen,Liu2025Phase}, a statistical framework originally developed to identify emergent large-scale organization and critical transitions in non-equilibrium systems across disciplines such as physics~\cite{liu2022renormalization,zhang2024eigen}, climate dynamics~\cite{ma2024frequency,ma2024increased}, neuroscience~\cite{chen2023leading}, and ecology~\cite{xie2025ecosystem,wang2024holistic}. Here, we adapt this framework to define an entropy metric that directly characterizes the disorder in SM dynamics.\par

We represent the system as $N$ spatial units, each with a state variable $X_{i}(t)$ at discrete time $t$. The evolving system state is expressed as a sequence of microstates $\boldsymbol{X}(t)$, which collectively form a normalized data matrix $\boldsymbol{A} \in \mathbb{R}^{N\times M}$:
\begin{equation}
    \boldsymbol{A}_{it} = \frac{X_i(t)}{\sqrt{C_0}}, \qquad \  C_0 = \sum_{t=1}^{M} \sum_{i=1}^{N} X_i^2(t),
\end{equation}
such that $\mathrm{Tr}(\boldsymbol{A}\boldsymbol{A}^\top) = 1$. In our analysis, $X_i(t) = \sqrt{Y_i(t)}$ where $Y_i(t)$ denotes SM or Pp.
We then decompose the system using singular value decomposition:
$
\boldsymbol{A} = \sum_{I=1}^N \sigma_I \boldsymbol{U}_I \otimes \boldsymbol{V}_I,
$
where $\boldsymbol{U}_I$ and $\boldsymbol{V}_I$ represent the spatial and temporal structures of the $I$-th eigen microstate, respectively, and $\sigma_I$ are the corresponding singular values satisfying $\sum_I \sigma_I^2=1$. $p_I = \sigma_I^2$ quantifies the fraction of system dynamics explained by each mode and can be interpreted as the probability of the system occupying the $I$-th eigen microstate. This representation effectively reduces the high-dimensional SM field to a set of dominant, independent spatial–temporal patterns with interpretable physical meaning. A detailed description of the EMT is provided in Supplementary Text.\par

Building on the probability distribution of eigen microstates ($\{p_I\}$), we define a Shannon-type entropy:
\begin{equation}
S_{\text{EM}} = - \sum_{I=1}^{N} p_I \ln p_I,
\label{eq:sem}
\end{equation}
which measures how evenly the system's dynamics are distributed among the eigen microstates. Low $S_{\text{EM}}$ indicates that variability is concentrated in one or a few coherent modes, reflecting an ordered system with strong organization. High $S_{\text{EM}}$ reflects a more uniform distribution with many competing modes, corresponding to disordered or noisy states with no dominant spatial–temporal pattern~\cite{Liu2025Phase}. In the theoretical limit of a perfectly ordered system (e.g., condensation into a single dominant eigen microstate), $p_1 \rightarrow 1$ and $S_{\text{EM}} \rightarrow 0$. Conversely, in a maximally disordered system (e.g., purely random system) with all eigen microstates contributing equally, $p_I \rightarrow 1/N$ and $S_{\text{EM}}\rightarrow \ln{N}$. Therefore, $S_{\text{EM}}$ serves as a sensitive and interpretable measure of disorder.\par

To pinpoint the specific structure driving systemic disorder, we decompose the total entropy into contributions from individual eigen microstates. The fractional contribution $c_I$ of the $I$-th eigen microstate is defined as:
\begin{equation}\label{eq:Ci}
c_I= s_I/{S_{\text{EM}}}, \qquad s_I = -p_I \ln(p_I),
\end{equation}
This decomposition serves as a diagnostic tool, allowing us to isolate the dominant modes responsible for the system's disorder. By ranking the eigen microstates in descending order of $c_I$, we identify the primary spatial patterns (or dominant modes) that govern the system's structural complexity. \par

\subsection{Local Facilitation Cellular Automaton model (LFCA)}

Traditional tipping point analyses often rely on one-dimensional stochastic models such as the Ornstein–Uhlenbeck (OU) process~\cite{smith2023reliability,boers2025destabilization}, which are inherently limited to a single state variable and cannot represent spatial interactions or multi-component feedbacks. However, many real-world eco-hydrological systems, including vegetation–soil moisture coupling, exhibit high-dimensional spatial organization and nonlinear local facilitation processes that can drive abrupt regime shifts. To illustrate how entropy behaves in such systems approaching an abrupt transition, we employ a LFCA model~\cite{kefi2007local}, which provides a generic yet powerful representation of spatially coupled, nonlinear dynamics with bifurcation-like behavior.\par

Following Ref.~\cite{tirabassi2023entropy}, the LFCA model evolves on a $100 \times 100$ lattice in discrete time. Each cell can occupy one of three possible states: $-1$, $0$, or $+1$. State transitions are probabilistic and depend both on the local neighborhood and on the global state fraction, capturing the interplay between short-range facilitation and system-wide conditions. The four possible transitions are: $+1 \;\rightleftharpoons\; 0 \;\rightleftharpoons\; -1$, with transition probabilities defined as
\begin{align}
w_{+|0} &= m, \\
w_{0|+} &= \delta \rho_+ + (1-\delta) q_{+|0} (b - c \rho_+), \\
w_{-|0} &= r + f q_{+|-}, \\
w_{0|-} &= d,
\end{align}
where $q_{A|B}$ denotes the fraction of neighboring units in state $A$ around a unit in state $B$, and $\rho_+$ is the global fraction of $+1$ units. The parameters $m$, $r$, $f$, $d$, $\delta$, and $c$ are constants (see Supplementary Table~3). The parameter $b$ serves as the control parameter, and as it decreases, the system undergoes an abrupt shift in its global state, allowing early-warning signals, such as rising entropy, to be observed prior to the transition. This rise in entropy emerges because the dominant pattern progressively weakens near the transition, allowing multiple competing modes to coexist, which increases the number of effective modes contributing to the dynamics.\par

\clearpage
\section*{Declarations}
\backmatter

\bmhead{Author contribution}
Y.X., T.L. and X.C. conceived and designed the study. Y.X. performed the simulations and data analyses with guidance from T.L. Y.X. and T.L. wrote the original draft of the manuscript. All authors discussed the results and contributed to manuscript revision.

\bmhead{Competing interests}

The authors declare no competing interests.

\bmhead{Supplementary information}

Supplementary information is available for this paper.

\bmhead{Data availability}
The GLDAS-NOAH data used here are publicly available at \url{https://disc.gsfc.nasa.gov/datasets/GLDAS_NOAH025_M_2.1/summary?keywords=GLDAS}. The ERA5 reanalysis data are publicly available at \url{https://cds.climate.copernicus.eu/datasets/reanalysis-era5-pressure-levels?tab=download}. The MERRA-2 reanalysis data are publicly available at \url{https://disc.gsfc.nasa.gov/datasets/M2TMNXLND_5.12.4/summary?keywords=M2CONXLND}. The ERA5-Land reanalysis data are publicly available at \url{https://cds.climate.copernicus.eu/datasets/reanalysis-era5-land-monthly-means?tab=download}. The CMIP6 data are publicly available at \url{https://aims2.llnl.gov/search}. The ONI data are publicly available at \url{https://origin.cpc.ncep.noaa.gov/products/analysis_monitoring/ensostuff/ONI_v5.php}. The MEI.v2 data are publicly available at \url{https://psl.noaa.gov/enso/mei/}.

\bmhead{Code availability}
The python code for analysis will be publicly available upon acceptance of this manuscript for publication.

\bmhead{Acknowledgements}
This work was supported by the National Key  Research and Development Program Program of China (Grant No. 2023YFE0109000) and the National Natural Science Foundation of China (Grant No. 12135003). This is ClimTip contribution \#; the ClimTip project has received funding from the European Union's Horizon Europe research and innovation programme under grant agreement No. 101137601. As Associated Partner the BNU has received funding from the Chinese Ministry for Science and Technology (MOST). The contents of this publication are the sole responsibility of the authors and do not necessarily reflect the opinion of the European Union or MOST.
\bibliography{ref} 
\end{document}